\documentclass[aos,MSNbibl,seceqn,nameyear,dvips]{arximspdf}

\doi{10.1214/14-AOS1241} 
\volume{42}
\issue{4}
\pubyear{2014}
\firstpage{1635}
\lastpage{1656}
\docsubty{FLA}

\makeatletter
\def\det{\operatorname{det}}
\def\tr{\operatorname{tr}}
\def\sup{\operatorname{sup}}
\def\diag{\operatorname{diag}}
\newtheorem{theor}{Theorem}[section]
\newtheorem{lemm}{Lemma}[section]
\newproclaim{exmpl}{Example}[section]
\newproclaim{rem}{Remark}[section]
\makeatother

\begin{document}
\begin{frontmatter}

\title{$E$-optimal designs for second-order response surface models}
\runtitle{$E$-optimal designs for second-order models}

\begin{aug}
\author[a]{\fnms{Holger}~\snm{Dette}\corref{}\ead[label=e1]{holger.dette@rub.de}\thanksref{t1,t2}}
\and
\author[b]{\fnms{Yuri}~\snm{Grigoriev}\thanksref{t1}\ead[label=e2]{yuri\_grigoriev@mail.ru}}
\runauthor{H. Dette and Y. Grigoriev}
\affiliation{Ruhr-Universit\"at Bochum and St.-Petersburg State
Electrotechnical University}
\address[a]{Fakult\"at f\"ur Mathematik\\
Ruhr-Universit\"at Bochum\\
44780 Bochum\\
Germany\\
\printead{e1}} 
\address[b]{Department of Computer Science\\
St.-Petersburg State Electrotechnical University\\
197376 St. Petersburg\\
Russia}
\end{aug}
\thankstext{t1}{Supported by the Deutsche Forschungsgemeinschaft (SFB 823: Statistik nichtlinearer dynamischer Prozesse, Teilprojekt C2).}
\thankstext{t2}{Partially supported by the National Institute of General Medical
Sciences of the National Institutes of Health under Award Number R01GM107639.}

\received{\smonth{3} \syear{2014}}
\revised{\smonth{5} \syear{2014}}

%
\begin{abstract}
$E$-optimal experimental designs for a second-order response surface model
with $k\geq1$ predictors are investigated. If the design space
is the \mbox{$k$-}dimensional unit cube, Galil and Kiefer
[\textit{J.~Statist. Plann. Inference} \textbf{1} (1977a) 121--132]
determined optimal
designs in a restricted class of designs
(defined by the multiplicity of the minimal eigenvalue) and stated
their universal optimality as a conjecture.
In this paper, we prove this claim and show that these designs are in fact
$E$-optimal in the class of all approximate designs. Moreover, if the
design space is the unit ball, $E$-optimal
designs have not been found so far and we also provide a complete
solution to this optimal design problem.

The main difficulty in the construction of $E$-optimal designs for the
second-order response surface model consists in the fact that for the
multiplicity of the minimum eigenvalue of
the ``optimal information matrix'' is larger than one (in contrast to
the case $k=1$) and as a consequence the corresponding optimality
criterion is
not differentiable at the optimal solution. These difficulties are
solved by considering nonlinear Chebyshev approximation problems,
which arise from a corresponding equivalence theorem. The extremal
polynomials which solve these Chebyshev problems
are constructed explicitly leading to a complete solution of the
corresponding $E$-optimal design problems.\looseness=-1
\end{abstract}

%
\end{frontmatter}

\section{Introduction}\label{sect1}

Response surface methodology has become a standard tool in the analysis
of experimental data. These models are used to study the influence of
several\vadjust{\goodbreak} input factors on a response variable by approximating complex
functional relationships by ``simple'' linear or quadratic multivariate
polynomial regression models, which are usually denoted as first or second-order
response surface models [see, e.g., \citet{myers2002}].
Numerous authors have worked on the construction of efficient and
optimal experimental designs for response surface models. For
first-order models, $2^k$ factorial and fractional factorial $2^{k-p}$
designs of resolution III are optimal with respect to the $D$-, $G$-
and $I$-optimality criteria [see \citet{andbormon2009}]. On the other
hand, for the second-order response surface model the situation is more
complicated and
intuitively reasonable designs with a ``simple'' structure such as
central composite designs are not optimal.

For this model, approximate designs in the sense of \citet{kiefer1974}
have been investigated by several authors, where the methodology and
optimal designs differ by the design space and optimality criterion
under consideration (typical a $k$-dimensional cube, ball or simplex).
$D$-optimal approximate designs for the second-order polynomial
regression model on the ball and cube have been determined explicitly
by \citeauthor{kiefer1959} (\citeyear{kiefer1959,kiefer1961}), \citet{kiewol1959},
\citet{kono1962},
\citet{farrell1967} [see also \citet{rafmys1988}, \citet
{limstu1988}
and \citet{detroe1997} who determined optimal product designs for
multivariate polynomial regression models
in more general situations].
In particular, it is shown that $D$-optimal designs on a ball are at
the same time rotatable designs. Considerably less attention has been
paid to other optimality criteria.
\citet{laptev1974}, \citet{denisov1976} and \citet{golikova1977}
investigated $A$- and $Q$-optimal designs numerically, \citet
{galil1977a} determined numerically rotatable optimal designs for the
second-order response surface model, while \citet{draheipuk2000} and
\citet{drappuk2003} investigated optimal design problems in
second-order mixture models.
On the other hand, the explicit determination of optimal designs in the
class of all approximate designs with respect to other criteria than
the $D$-criterion seems to be a very hard problem, which has only been
solved in rare circumstances.

In this paper, we study $E$-optimal designs for the second-order
response surface models on the $k$-dimensional cube and ball.
Among Kiefer`s $\Phi_p$-criteria [see \citet{kiefer1974}] the
$E$-optimality criterion is not differentiable
if the multiplicity of the minimum eigenvalue of the information matrix
of the optimal design is larger than $1$. This property makes the determination
of $E$-optimal designs to an extremely hard and challenging problem. In fact,
an analytical construction of \mbox{$E$-}optimal designs for linear regression
models is very difficult and has only been achieved in the one-dimensional
case for a limited number of linear and nonlinear models [see
\citet{melas1982}, \citet{dette1993}, \citet
{pukstu1993}, \citet{dethai1994},
among others].
For models with more than one predictor, results can only be found
sporadically in the literature.
For example, \citet{cheng1987} and \citet{detstu1993} identified
$E$-optimal spring balance and chemical balance weighing designs.
\citet{galil1977b} considered the second-order response surface model
on the cube with $k$ predictors and determined the $E$-optimal designs
in the class of all designs, for which the corresponding information
matrix has a minimum eigenvalue
of multiplicity ${k(k+1)}/{2}$. However, to our best knowledge,
the answer to the question, if these
designs are in fact \mbox{$E$-}optimal in the class of all designs is still
open. For the ball, the situation is even worse, and only $E$-optimal
designs in
the class of all rotatable designs are available [see, e.g., \citet{galil1977a}]. These designs are in fact not globally optimal and the
determination of $E$-optimal designs for the
second-order response surface model on the ball is an open and
challenging problem.

The goal of the present paper is to provide complete answers to these
questions and to characterize the structure and properties of
$E$-optimal designs for the second-order response surface model. Our
approach relies on a specific duality result
for $E$-optimal designs, which relates the optimal design problem to a
nonlinear Chebyshev approximation problem
[see \citeauthor{melas1982} (\citeyear{melas1982,melas2006}) or \citet{pukelsheim2006}].
In the
dual problem, one has to determine a nonnegative polynomial
with minimal $\sup$-norm in a specific class of nonnegative
(multivariate) polynomials, that is,
%
%
\begin{equation}
\label{extrpol} {\mathcal P} = \bigl\{ f^T(x) Z f(x)| \operatorname{trace}
(Z)=1; Z\ge 0 \bigr\},
\end{equation}
where $x$ denotes the $k$-dimensional predictor, $f(x)$ is the vector
of regression functions
in the second-order response surface model and $Z$ is a nonnegative
definite matrix of appropriate dimension.
This Chebyshev approximation problem is nonlinear and, therefore,
extremely hard to solve explicitly. For the solution of the $E$-optimal
design problem, this ``optimal'' polynomial, which is called extremal
polynomial throughout this paper,
will be constructed explicitly in Sections~\ref{sect3} and~\ref{sect4} if the design space is the cube
and ball, respectively. As a consequence, we are able to provide a
complete solution of these $E$-optimal
design problems. In general, there exist several $E$-optimal designs
which usually have a large number
of support points. For this reason, particular attention is paid to the
problem of constructing $E$-optimal
designs with a small number of support points.

\section{Optimal designs for response surface models}\label{sect2}

We consider the common linear regression model of the form
%
%
\begin{equation}
\label{unconditionedregression} \mathbb{E}(Y|x)=f^T(x) \theta,
\end{equation}
where $Y$ denotes the (one-dimensional) response and the explanatory
variable $x$ varies in a compact design space, say $\mathcal{X}
\subset\mathbb{R}^k$. In (\ref{unconditionedregression}), the vector
$f(x)=(f_1(x),\ldots,f_m(x))^T \in\mathbb{R}^{m}$ is the vector of
regression functions and
$\theta= (\theta_1, \ldots, \theta_m) ^T\in\mathbb{R}^{m} $
denotes a vector of unknown parameters.
We assume that $N$ independent observations are available according to
the model (\ref{unconditionedregression}) where
at each experimental condition $x$ the response $y$
is a realization of a normal distributed random variable $Y$ with
expectation given by (\ref{unconditionedregression})
and (constant) variance $\sigma^2 > 0$. An approximate design in the
sense of \citet{kiefer1974} is defined as probability measure on the
design space $\mathcal{X}$ with finite support.
The support points, say $x_{(1)},\ldots, x_{(n)}$, of an approximate
design $\xi$ define the locations where observations
are taken, while the weights give the corresponding relative
proportions of total observations to be taken at these points.
If the design $\xi$ has masses $\omega_i>0 $ at the different points
$x_{(i)}$  $ (i =
1,\ldots, n)$ and $N$ observations can be made by the experimenter,
the quantities
$\omega_i N$ are rounded to integers, say $N_i$, satisfying $\sum^n_{i=1} N_i =N$, and
the experimenter takes $N_i$ observations at each location $x_{(i)}$
$(i=1,\ldots, n)$.
The information matrix of an approximate design $\xi$ is defined by
%
%
\begin{equation}
\label{11a} M(\xi) = \int_{\mathcal X} f(x) f^T(x) \,d
\xi(x) \in\mathbb{R}^{m \times m},
\end{equation}
and it is well known [see \citet{jennrich1969}] that under appropriate
assumptions of
regularity [in particular $\det(M(\xi)) > 0$ and $\lim_{N_i, N \to
\infty} N_i / N = \omega_i>0 $; $i=1,\ldots, n$] the covariance
matrix of the least squares estimator is approximately
given by $\sigma^2  M^{-1}(\xi)/N$, where $N$ denotes
the total sample size.

Optimal designs maximize an appropriate statistical meaningful
functional, say~$\Phi$, of the information matrix.
Among the numerous criteria which have been proposed in the
literature for this purpose [see \citet{silvey1980}, \citet
{pazman1986}
or \citet{pukelsheim2006} among others],
we consider in this paper the $E$-optimality criterion
%
%
\begin{equation}
\label{eopt} \Phi_{-\infty}(\xi) = \lambda_{\min} \bigl( M(\xi)
\bigr).
\end{equation}
This criterion arises as a special case of Kiefer's $\Phi
_p$-optimality criteria, which are defined for $p\in(-\infty, 1]$ as
%
%
\begin{equation}
\label{Kiefertracecriterion411} \Phi_p(M)= \bigl[m^{-1}\tr
\bigl(M^{p}(\xi)\bigr) \bigr]^{1/p}= \Biggl(m^{-1}
\sum_{i=1}^m\lambda_i^{p}
\bigl(M (\xi)\bigr) \Biggr)^{1/p},
\end{equation}
that is $ \Phi_{-\infty} (\xi) = \lim_{p \to-\infty}\Phi
_p(\varepsilon)$ [see \citet{kiefer1974}].
In equation (\ref{Kiefertracecriterion411}), the quantities
$\lambda_1 ( M(\xi)), \ldots, \lambda_m ( M(\xi))$
denote the eigenvalues of the information matrix $M(\xi)$
and $ \lambda_{\min}
( M(\xi)) $ its corresponding minimum eigenvalue.
In contrast to the $\Phi_p$-criteria with $p \in(-\infty, 1]$ the
$E$-optimality criterion is not differentiable
if the multiplicity of the minimum eigenvalue of the matrix $M(\xi)$
is larger than $1$ and this property makes the determination
of $E$-optimal designs to an extremely hard problem. In fact,
$E$-optimal designs have been determined
for a limited number of linear and nonlinear regression models [see the
references cited in the \hyperref[sect1]{Introduction}]. An important tool for the determination
of $E$-optimal designs is the following equivalence theorem which has
been proved by several authors [see \citet{melas1982} or
\citet{pukelsheim2006}, e.g.].

%
\begin{theor} \label{theoremE-optimalityconditionofMelas}
Let $\xi^*$ denote a design and $\lambda_{\min}(M(\xi^*))$ the
minimum eigenvalue of the information matrix $M(\xi^*)$ with
multiplicity $s$. The design
$\xi^*$ is \mbox{$E$-}optimal if and only if there exist
orthonormal eigenvectors $q_0, \ldots,q_{s-1}$ of the matrix $M(\xi
^*)$ corresponding to $\lambda_{\min} (M(\xi^*)) $
and nonnegative weights $w_0, \ldots, w_{s-1}$ with sum $1$ such that
the ``extremal polynomial''
\begin{eqnarray*}
d(x,\xi) &=& f^T(x) (q_0,\ldots, q_{s-1}) \diag (w_0,\ldots, w_{s-1} ) (q_0,\ldots,q_{s-1})^T f(x)
\\
&=& \sum^{s-1}_{i=0}
w_i \bigl(f^T(x)q_i\bigr)^2
\end{eqnarray*}
satisfies for all $x \in\mathcal{X}$ the inequality
%
%
\begin{equation}
\label{A-matrixofMelassequalandmore2} d(x,\xi) \leq\lambda_{{\min}}\bigl(M\bigl(\xi^*\bigr)\bigr).
\end{equation}
Moreover, the maximum on the left-hand side of (\ref
{A-matrixofMelassequalandmore2})
is attained at the support points of the $E$-optimal design $\xi^*$.
\end{theor}

%
%
\begin{rem}
It follows from general equivalence theory developed in convex
design theory [see \citet{pukelsheim2006}] that there exists a duality
between the $E$-optimal design problem and a nonlinear Chebyshev
approximation problem, that is,
%
%
\begin{equation}
\label{dual} \max_\xi\lambda_{\min} \bigl(M(\xi)
\bigr) = \min_{P_Z \in\mathcal{P}} \max_{x \in\mathcal{X}}
\bigl|P_Z (x)\bigr|,
\end{equation}
where $\mathcal{P}= \{ P_Z (x) = f^T(x) Z f (x) | Z \in\mathbb
{R}^{m \times m},  Z \geq0,  \operatorname{trace}(Z)=1 \}$ denotes a subset
of the nonnegative ``polynomials.'' In fact, if there is equality in
(\ref{dual}) for a pair $(\xi^*, Z^*)$, then $\xi^*$ is an
$E$-optimal design and $P_{Z^*}$ a solution of the nonlinear Chebyshev
approximation problem. This explains the name ``extremal polynomial''
in Theorem~\ref{theoremE-optimalityconditionofMelas}.
\end{rem}

The second-order response surface model with a $k$-dimensional
predictor appears as a special case of model (\ref
{unconditionedregression}), that is,
%
%
\begin{equation}
\label{quadsurf} \mathbb{E}[Y|x] = \sum_{\| \alpha\|_1 =0}^2
\theta_\alpha x^\alpha,
\end{equation}
where $\alpha=(\alpha_1,\ldots, \alpha_k)^T \in\{0,1,2 \}^k$ is a
multiindex $x^\alpha= x^{\alpha_1} \cdots x^{\alpha_k}$ and
$ \| \alpha\|_1=\alpha_1+\cdots+\alpha_k$.
In this case, the corresponding vector of regression function in the
general linear model
(\ref{unconditionedregression}) is given by
%
%
\begin{equation}
f (x)= \bigl(1, x_1^2,\ldots, x_k^2,
x_1,\ldots, x_k, x_1x_2,\ldots,
x_{k-1}x_k\bigr) \in\mathbb{R}^m, \label{basicfunctionofquadraticmodel}
\end{equation}
where\vspace*{1pt} $m=\frac{(k+1)(k+2)}{2}$, $x=(x_1,\ldots, x_k)^T$.
In the following section, we consider optimal designs for the
second-order regression model (\ref{quadsurf}), where the design
spaces are the unit ball
with respect to the maximum norm $\|\cdot\|_{\infty}$ and the
Euclidean norm $\|\cdot\|_{2}$, that is,
%
%
\begin{eqnarray}\label{cubeandball}
{\mathcal X} &=& \mathcal{B}_\infty(1):=\bigl\{x\in
\mathbb{R}^k | \|x\| _{\infty} \leq1 \bigr\},
\nonumber\\[-8pt]\\[-8pt]
{\mathcal X}&=& \mathcal{B}_2 (1):= \bigl\{ x\in\mathbb{R}^k | \|x
\|_{2} \leq 1 \bigr\}.\nonumber
\end{eqnarray}
It turns out that designs with certain symmetry properties play a particular
role for the construction of $E$-optimal designs. Throughout this paper,
we call a design \textit{symmetric} if for
any $(\alpha_1,\ldots, \alpha_k) \in\{ 0,1,2\}^k$ with $\| \alpha
\|_1 = |\alpha_1| + \cdots+ |\alpha_k| \leq2$
the moments
\[
\int_{\mathcal X} x_1^{\alpha_1}, \ldots,
x_k^{\alpha_k}\xi(dx)
\]
are invariant with respect to all permutations of $\alpha_1,\ldots,
\alpha_k$
and vanish if there is at least one odd index among $\alpha_1, \ldots,\alpha_k$.
In the following discussion, let $I_\ell\in\mathbb{R}^{\ell\times
\ell} $ denote the identity matrix and $1_\ell=(1,\ldots,1)^T \in
\mathbb{R}^\ell$ denotes the vector
with all elements equal to $1$, then a straightforward calculation shows
that the information matrix of a symmetric design in model (\ref
{quadsurf}) is of the form
%
%
\begin{eqnarray}\label{informationmatrixforquadraticmodel}
M(\xi) &=& \int_\mathcal{X}f(x)f^T(x)\xi(dx)
\nonumber\\ \\[-16pt]
&=& \pmatrix{ 1 & a1^T_k & 0 & 0
\cr
a1_k
& H & 0 & 0
\cr
0 & 0 & aI_k & 0
\cr
0 & 0 & 0 & bI_{({k(k-1)})/{2}}}
\in\mathbb{R}^{m \times m},\nonumber
\end{eqnarray}
where $m= \frac{(k+1)(k+2)}{2}$, $H=H(c;b) = (c-b) I_k + b 1_k1_k^T\in
\mathbb{R}^{k\times k} $ denotes a circulant matrix with diagonal and
off-diagonal elements $c$ and $b$, respectively, and the entries
$a,b$ and $c$ in (\ref{informationmatrixforquadraticmodel}) are given by
%
%
\begin{equation}
\label{momentsofabandc} a=\int_\mathcal{X}x_1^2
\xi(dx), \qquad b=\int_\mathcal {X}x_1^2x_2^2
\xi(dx), \qquad c=\int_\mathcal{X}x_1^4
\xi(dx).
\end{equation}

Designs with information matrix of the form (\ref
{informationmatrixforquadraticmodel}) will serve as candidates for
$E$-optimal designs. Consider, for example, the case
$k=1$, where model~(\ref{quadsurf}) reduces to the well-known
one-dimensional quadratic regression model \mbox{$ \theta_0 +\theta_1x^2 +\theta_2x$}. If the designs space is given
by ${\mathcal X} = [-1, 1]$ and the design $\xi$
puts masses $1/5$, $1/5$ and $3/5$ at the points $-1$, $1$ and $0$,
respectively, the corresponding
information matrix is given by
\[
M\bigl(\xi^*\bigr)= \pmatrix{ 1& \frac{2}{5} & 0
\vspace*{3pt}\cr
\frac{2}{5} &
\frac{2}{5} & 0
\vspace*{3pt}\cr
0 & 0 & \frac{2}{5}}.
\]
It was shown by
\citet{kiefer1974}
that this design is in fact $E$-optimal for the univariate quadratic
regression model and the minimum
eigenvalue $\lambda_{\min}=\frac{1}{5} $ has multiplicity $ s=1$.
For a similar statement in the univariate polynomial regression model
of degree $d \ge2$, see
\citet{pukstu1993}.

However, in the case $k\ge2$, the multiplicity of the minimum eigenvalue
of the matrix (\ref{informationmatrixforquadraticmodel}) is larger
than $1$ and as consequence the corresponding optimality criterion is
not differentiable at the matrix $M(\xi)$ given by (\ref
{informationmatrixforquadraticmodel}). This makes the determination of
$E$-optimal designs substantially more difficult. For example,
\citet{galil1977b} determined the $E$-optimal design on the cube
$\mathcal
{B}_\infty(1)$ in the subclass of all designs with information matrix
of the form (\ref{informationmatrixforquadraticmodel}), where its
minimum eigenvalue has multiplicity $\frac{k(k+1)}{2}$ (these
calculations will be briefly presented at the beginning of the
following section). To our best knowledge, the question, if the
solution obtained by these authors in the restricted class yields in
fact an $E$-optimal design for the
second-order response surface model in the class of all approximate
designs on the cube, has not been answered. Moreover, the $E$-optimal
design problem for
second-order regression models seems to be completely unsolved if the
design space is given by the unit ball $\mathcal{B}_2(1)$.

In the following two sections, we will present a complete solution to
these problems.
For this purpose, we proceed in the following sections in two steps:
\begin{longlist}[(II)]
\item[(I)] In a first step, a candidate for the $E$-optimal design in
the class of all designs with information matrix of the form (\ref
{informationmatrixforquadraticmodel}) is identified. If the design
space is given by the cube, our arguments coincide with those of
\citet{galil1977b} and are presented here for the sake of completeness.
\item[(II)] In a second step, the $E$-optimality of the candidate
design found by \citet{galil1977b} is proved by an application of
Theorem~\ref{theoremE-optimalityconditionofMelas}.
This requires the determination of an appropriate basis of the
eigenspace corresponding to the minimum eigenvalue of $M(\xi)$ and the
construction of the corresponding extremal polynomial
in (\ref{dual}).
\end{longlist}
The $E$-optimal designs for the second-order response surface model
will be identified in terms of the masses that they assign to specific
sets which depend on the design space under consideration.
Because in many applications it is desirable to obtain optimal designs
with a minimal number of support points, we
add a third step if the design space is the cube, that is,
\begin{longlist}[(III)]
\item[(III)] Identification of designs with a minimal number of
support points.
\end{longlist}

\section{$E$-optimal designs on the cube}\label{sect3}

In this section, we consider the second-order response surface model
(\ref{unconditionedregression}) on the design space $\mathcal
{X}=\mathcal{B}_\infty(1)=[-1,1]^k$.
We start with a determination of a ``good'' candidate for an
$E$-optimal symmetric design. Our arguments are similar to those given
in \citet{galil1977b}
and presented here for the sake of completeness (note that these
authors only identified the candidate design
and in the following we will prove its optimality in the class of all
approximate designs). Observing the representation of the corresponding
information matrix
(\ref{informationmatrixforquadraticmodel}) the eigenvalues of the matrix
$M(\xi)$ are given by $a$, $b$, and the eigenvalues by its upper
$(k+1) \times(k+1)$ block,
%
%
\begin{equation}
\label{matrixA} M_{11}(\xi) =\pmatrix{ 1_k &
a1^T_k
\vspace*{3pt}\cr
a1_k & H},
\end{equation}
where $H=H(c;b)=(c-b)I_k + b 1_k 1^T_k$. Define
$D= [1-c-(k-1)b]^2+ 4ka^2>0$, then all eigenvalues of the information
matrix of a symmetric $E$-optimal design are given by
%
%
\begin{eqnarray}\label{eigenvaluesab}
\lambda_{0} &=& \frac{1+c+(k-1)b + \sqrt{D}}{2}, \qquad
\lambda_{1} = \frac{1+c+(k-1)b - \sqrt{D}}{2}, \nonumber
\\
\lambda_2 &=& \cdots=\lambda_{k}=c-b, \qquad
\lambda_{k+1} = \cdots =\lambda_{2k}=a,
\\
\qquad\lambda_{2k+1} &=& \cdots=\lambda_m=b.\nonumber
\end{eqnarray}
Note that $\lambda_0>\lambda_1$ and that $\lambda_1$ and $\lambda_2$
are increasing functions of $c$.
Observing the identity
\[
\det M(\xi) = a^kb^{k(k-1)/2} (c-b)^{k-1}
\bigl[c+(k-1)b-ka^2\bigr]>0 %
\]
it is easy to see that the entries of a nonsingular matrix of the form
(\ref{informationmatrixforquadraticmodel}) satisfy the inequalities
%
%
\begin{equation}
\label{inequalityabc} 1>\ge a\geq c> b> 0, \qquad c+b (k-1)> ka^2.
\end{equation}
Therefore, we obtain $c=a$ and the problem of maximizing
the minimum eigenvalue of $M(\xi)$ reduces to the maximization of
%
%
\begin{eqnarray}\label{lammin}
\lambda_{\min}\bigl(M(\xi)\bigr) &=& \min \biggl\{
\frac{1+a+(k-1)b- \sqrt{D}}{2}, a-b, a,b \biggr\}
\nonumber\\[-8pt]\\[-8pt]
&=& \min \biggl\{ \frac{1+a+(k-1)b- \sqrt{D}}{2}, a-b,b\biggr\},\nonumber
\end{eqnarray}
where the constant $D$ is now represented as $D=[1-a-(k-1)b]^2+4ka^2$
and the second equality in (\ref{lammin}) follows from $0 < a-b < a$
[see (\ref{inequalityabc})].
We will now construct a candidate for the $E$-optimal design.
Motivated by the solution of similar maximin problems, we suppose for
this purpose that
\[
\lambda_1 = \frac{1+a+(k-1)b-\sqrt{D}}{2} = a-b=b, %
\]
which gives
$a=\frac{2}{5}$, $b=\frac{1}{5}$ as a unique (nontrivial) solution.
This yields for the eigenvalues of the matrix $M(\xi)$
%
%
\begin{eqnarray}\label{optimaleigenvalues}
\lambda_0&=&1+\frac{k}{5},\qquad
\lambda_1=\cdots=\lambda_k=\frac{1}{5},
\nonumber\\[-8pt]\\[-8pt]
\lambda_{2k+1}&=&\cdots=\lambda_{(k(k-1))/2}=\frac{1}{5}, \qquad
\lambda_{k+1}=\cdots=\lambda_{2k}=\frac{2}{5},\nonumber
\end{eqnarray}
where\vspace*{1pt} the corresponding multiplicities of $\lambda_0, \lambda_1,
\lambda_{k+1}$ are given by
$1, \frac{k(k+1)}{2}$ and $k$, respectively.
Hence, we obtain as a candidate for an $E$-optimal information matrix
the matrix $M(\xi^*)$ in (\ref{informationmatrixforquadraticmodel})
with $a=c= \frac{2}{5},  b=\frac{1}{5}$, where the minimum
eigenvalue is given by $\lambda_{\min}(M(\xi^*))=\frac{1}{5}$.
This means that the information matrix under consideration has a
minimal eigenvalue with
multiplicity $\frac{k(k+1)}{2} \geq3$ whenever $k \geq2$.
The following result gives an answer to the question if the determined
values for $a$ and $b$ yield in fact to an $E$-optimal information matrix.

%
%
\begin{theor}
\label{theorem_e_optimalityoncubeof_dette_grigoriev}
Any design $\xi^*$ with an information matrix $M(\xi^*)$ of the form~(\ref{informationmatrixforquadraticmodel}) and $a=c={2\over5}$
$b={1\over5}$
is $E$-optimal for the second-order response surface model (\ref
{quadsurf}) on the $k$-dimensional unit cube.
In particular, Theorem~\ref{theoremE-optimalityconditionofMelas} holds with
%
%
\begin{equation}
\label{extremalMelaspolynomfactor5} d(x,\varepsilon)= \frac{1}{5}
\Biggl(1-\frac{4}{k}\sum_{i=1}^kx_i^2
\bigl(1-x_i^2\bigr) \Biggr).
\end{equation}
\end{theor}

The proof of Theorem~\ref{theorem_e_optimalityoncubeof_dette_grigoriev} is
complicated and deferred to Appendix~\ref{61}.
Note that in contrast to the
$D$-optimality criterion the optimal values for $a$ and $b$ do not
depend on a dimension of the design space. This fact has been
independently observed by \citet{denisov1976} and \citet{galil1977b},
who identified the correct $E$-optimal information matrix but did not
prove its optimality.

In the next step, we determine designs with corresponding information
matrix specified in Theorem~\ref{theorem_e_optimalityoncubeof_dette_grigoriev}. For this purpose, we
call a point $x \in\mathbb{R}^k$ a barycenter of depth $0 \leq j \leq
k$ if $j$ coordinates are equal to $0$ and the remaining $k-j$
coordinates are equal to $\pm1$ [see \citet{galil1977b}].
The set of all barycenters of depth~$r$ is denoted $E_r$ and for its
cardinality we introduce the symbol
%
%
\begin{equation}
\label{numberpointsofEr} n_r:=|E_r|= \pmatrix{k \cr r}
2^{k-r}, \qquad r=0,1,\ldots, k.
\end{equation}
It was shown by \citet{kiefer1960} and \citet{farrell1967}
that the
support of every $\Phi_p$-optimal design for the
second-order response surface
model on the cube is a subset
of the set
%
%
\begin{equation}
\label{eset} E= \bigcup^k_{j=0}E_j.
\end{equation}
Moreover, there always exists a symmetric optimal design. Throughout
this section,
we will describe these symmetric designs on the cube in terms of the
\mbox{$(k+1)$-}dimensional vector $\xi=(\xi_0,\ldots, \xi_k)^T$, where
$\xi_i$ represents the mass
assigned by the design to the set $E_i$ of barycenters of depth $i$,
that is $\xi_i=\xi(E_i)$ ($i=0,\ldots, k$).
It turns out that there always exists an $E$-optimal design supported
at most three sets $E_i$.
For this purpose, we define for integers $0 \le r_1 < r_2 < r_3 \le k$
the matrix
\[
A_{r_1,r_2,r_3} =\pmatrix{ 1 & 1 & 1
\vspace*{5pt}\cr
\dfrac{k-r_1} {k} & \dfrac{k-r_2} {k} & \dfrac{k-r_3} {k}
\vspace*{7pt}\cr
\dfrac{k-r_1} {k} \dfrac{k-r_1-1} {k-1} &
\dfrac{k-r_2} {k} \dfrac{k-r_2-1} {k-1} & \dfrac{k-r_3} {k}\dfrac{k-r_3-1} {k-1}}.
\]

%
%
\begin{lemm} \label{design}
There exists integers $0 \le r_1 < r_2 < r_3 \le k$ such that the
system of linear equations
%
%
\begin{equation}
\label{aequat} A_{r_1,r_2,r_3} \xi= \bigl( 1, \tfrac{2}{5}, \tfrac{1}{5} \bigr)^T
\end{equation}
has a unique solution $\xi^* =(\xi_1^*,\xi_2^*,\xi_3^*)^T$
satisfying $\xi_i ^* \geq0$, $\sum_{i=1}^3 \xi_i^* =1$. Any design
with masses
%
%
\begin{equation}
\label{optmass} \xi(E_{r_i}) = \xi_i^*,\qquad i=1,2,3,
\end{equation}
is $E$-optimal for the second-order response surface model (\ref{quadsurf}).
\end{lemm}

\begin{pf}
Let $\xi$ denote a symmetric design and
note that the moments in the matrix $M(\xi)$ defined in (\ref
{informationmatrixforquadraticmodel}) have the representation
%
%
\begin{equation}
\label{expansionsofaandb} 1=\sum_{r=0}^k
\xi_r, \qquad a=\sum_{r=0}^{k-1}a_r
\xi_r, \qquad b=\sum_{r=0}^{k-2}b_r
\xi_r,
\end{equation}
where $ \xi_r= \xi(E_r) $ is the measure of the set $E_r$ of
barycenters of depth $r$ and
%
%
\begin{eqnarray}\label{expansioncoefficientsofaandb}
a_r &:=& \pmatrix{ k-1 \cr r } 2^{k-r}, \qquad r\in
\{0,\ldots,k-1\},
\nonumber\\[-8pt]\\[-8pt]
b_r &:=& \pmatrix{ k-2 \cr r } 2^{k-r},\qquad r\in\{ 0,\ldots,k-2 \}.\nonumber
\end{eqnarray}
By (\ref{expansionsofaandb}) and a remark on page~124 of \citet
{galil1977b}, there exist
symmetric design $\xi$ and three sets $E_{r_1}, E_{r_2}$ and $E_{r_3}$
such that
(\ref{expansionsofaandb}) is satisfied for $a={2\over5} $ and
$b={1\over5}$.
A simple calculation shows that in this case the system of equations in
(\ref{expansionsofaandb})
is equivalent to (\ref{aequat}), which has a unique solution because
$ \det(A) = \frac{(r_1-r_2)(r_1-r_3)(r_2-r_3)}{k^2(k-1)} \neq0$.
\end{pf}

It should be noted that not any solution of (\ref{aequat}) will yield
a vector of admissible weights $(\xi_{r_1},\xi_{r_2}, \xi_{r_3}) =
(\xi(E_{r_1}), \xi(E_{r_2}), \xi(E_{r_3})) $ (some components could
be negative).
Moreover, in general there exist many triples $(r_1,r_2,r_3)$, such
that the system (\ref{aequat}) has a solution with nonnegative
components and any such triple yields to at least one symmetric
$E$-optimal design.
For example, if $(r_1,r_2,r_3)$ is such a triple with corresponding
solution $(\xi(E_{r_1}), \xi(E_{r_2}), \xi(E_{r_3}))$ of (\ref{aequat}), then a design~$\xi$ which assigns masses
\[
\omega_{r_i,j} = \xi\bigl(\{ x_{r_{i,j}} \}\bigr) =
{\xi(E_{r_i}) \over{
n_{r_i} }}; \qquad j=1,\ldots,n_{r_i};  i=1,2,3;
\]
to all points $ x_{(r_i,1)} \cdots x_{(r_i,n_{r_i})} \in E_{r_i}$ is an
$E$-optimal design for the second-order response surface model
(\ref{quadsurf}) on the unit cube $[-1,1]^k$, where $n_j ={ k \choose
j}2^{k-j} $\vspace*{1pt} denotes the number of elements of the set $E_j$
($j=0,\ldots, k$).
The number of support points of such a design is given by
\[
N(r_1,r_2,r_3) = \sum
^3_{i=1} \pmatrix{k \cr r_i}
2^{k-r_i} %
\]
and usually rather large. For this reason, it is of interest to find
designs with a minimal number of support points
[see \citet{farrell1967} or \citet{pesotchinsky1975}].
A reasonable approach to this problem is to look for \mbox{$E$-}optimal
designs which are supported at only \emph{two} sets of barycenters,
say $E_{r_1}$ and~$E_{r_2}$. Because it can easily be shown that for a
triple $(r_1,r_2,r_3)$ with an admissible solution of (\ref{aequat}) the
weights $\xi(E_{r_i})$ are given by
%
%
\begin{equation}
\label{basicequationforEoptimaldesigns} \xi(E_{r_1})=\frac{1}{5}\cdot\frac{2k^2+k-
3k({r_2}+{r_3})+5{r_2}{r_3}}{({r_2}-{r_1})({r_3}-{r_1})},
\qquad i=1,2,3,
\end{equation}
it follows that symmetric $E$-optimal designs supported at only two
sets of barycenters can be obtained from the
\emph{Diophantine equations}
%
%
\begin{equation}
\label{diofantequation} 2k^2+k- 3k(s+t)+5st=0
\end{equation}
for $s,t =0,\ldots, k$. These equations have been solved numerically
by \citet{galil1977b} if $k\le25$ (see Table~1 in this
reference).
It should be pointed here that there does not always exist a solution
of (\ref{diofantequation})
(e.g., for $k=2,6$ or~$8$). Moreover, in general it is not clear that a
solution of (\ref{diofantequation})
necessarily yields to an $E$-optimal design with a minimal number of
support points. For this reason, we display
in Table~\ref{table31} the $E$-optimal symmetric designs with a
minimal number of support points for second-order response surface models
with $k\le24$ predictors. For example, if $k=5$, the design with a
minimal number of support points in only two sets has $N(2,5)=81$
support points in the set $E_2$ and $E_5$
[see \citet{galil1977b}], while the design with the minimal number of
$N(0,3,5)=73$ support points in the sets $E_0$, $E_3$ and~$E_5$.

%
%
\begin{table}
\caption{Symmetric $E$-optimal designs with a minimal number
of support points for second-order response surface models with $k \le 24$ predictors}\label{table31}
\tabcolsep=0pt
\begin{tabular*}{\tablewidth}{@{\extracolsep{\fill}}@{}lccccccccc@{}}
\hline
$\bolds{k}$ & $\bolds{(r_1,r_2,r_3)}$ &$\bolds{\xi(E_{r_1})}$ &$\bolds{\xi(E_{r_2})}$ &$\bolds{\xi(E_{r_3})}$
            & $\bolds{k}$ & $\bolds{(r_1,r_2,r_3)}$ &$\bolds{\xi(E_{r_1})}$ &$\bolds{\xi(E_{r_2})}$ &$\bolds{\xi(E_{r_3})}$
\\
\hline
\phantom{0}1 &$(0,1, \mbox{--})$ &$\frac{2}{5}$ &$\frac{3}{5}$ &-- &13 &$(0,9, \mbox{--})$&$\frac{2}{15}$ &$\frac{13}{15}$ &-- \\[4pt]
\phantom{0}2 &$(0,1, 2 )$ &$\frac{1}{5}$ &$\frac{2}{5}$ &$\frac{2}{5}$ &14&$(0,9, 14)$ &$\frac{25}{225}$ &$\frac{182}{225}$ &$\frac{18}{225}$\\[4pt]
\phantom{0}3 &$(\mbox{--},1, 3 )$ & -- &$\frac{3}{5}$ &$\frac{2}{5}$ &15 &$(0,10, 15)$&$\frac{3}{25}$ &$\frac{21}{25}$ &$\frac{1}{25}$ \\[4pt]
\phantom{0}4 &$(0,3, \mbox{--})$ &$\frac{1}{5}$ &$\frac{4}{5}$ &-- &16 &$(0,11, \mbox{--})$&$\frac{7}{55}$ &$\frac{48}{55}$ &-- \\[4pt]
\phantom{0}5 &$(0,3, 5 )$ &$\frac{2}{15}$ &$\frac{10}{15}$ &$\frac{3}{15}$ &17&$(0,11, 17)$ &$\frac{18}{165}$ &$\frac{136}{165}$ &$\frac{11}{165}$\\[4pt]
\phantom{0}6 &$(0,4, 6 )$ &$\frac{3}{20}$ &$\frac{15}{20}$ &$\frac{2}{20}$ &18&$(0,12, 18)$ &$\frac{7}{60}$ &$\frac{51}{60}$ &$\frac{2}{60}$ \\[4pt]
\phantom{0}7 &$(0,5, \mbox{--})$ &$\frac{4}{25}$ &$\frac{21}{25}$ &-- &19 &$(0,13, \mbox{--})$ &$\frac{8}{65}$ &$\frac{57}{65}$ &-- \\[4pt]
\phantom{0}8 &$(0,5, 8 )$ &$\frac{9}{75}$ &$\frac{56}{75}$ &$\frac{10}{75}$ &20&$(0,13, 20 )$ &$\frac{49}{455}$ &$\frac{380}{455}$ &$\frac{26}{455}$ \\[4pt]
\phantom{0}9 &$(0,6, 9 )$ &$\frac{2}{15}$ &$\frac{12}{15}$ &$\frac{1}{15}$ &21&$(0,14, 21 )$ &$\frac{4}{35}$ &$\frac{30}{35}$ &$\frac{1}{35}$ \\[4pt]
10 &$(0,7, \mbox{--})$ &$\frac{1}{7}$ &$\frac{6}{7}$ &-- &22 &$(0,15, \mbox{--})$&$\frac{3}{25}$ &$\frac{22}{25}$ &-- \\[4pt]
11 &$(0,7, 11)$ &$\frac{8}{70}$ &$\frac{55}{70}$ &$\frac{7}{70}$ &23&$(0,15, 23 )$ &$\frac{32}{300}$ &$\frac{253}{300}$ &$\frac{15}{300}$ \\[4pt]
12 &$(0,8, 12)$ &$\frac{5}{40}$ &$\frac{33}{40}$ &$\frac{2}{40}$ &24&$(0,16, 24 )$ &$\frac{9}{80}$ &$\frac{69}{80}$ &$\frac{2}{80}$\\
\hline
\end{tabular*}
\end{table}

%
%
\begin{table}[b]
\tabcolsep=4.4pt
\caption{Conjecture for the structure of $E$-optimal designs
with a minimal number of support points
for second-order response surface models with $k=1,2$ and $k \ge4 $
predictors, where $k=3q+l$ and $s=2q+l$ and $l=0,\pm1$}\label{table32}
\begin{tabular*}{\tablewidth}{@{\extracolsep{\fill}}@{}lccccccccc@{}}
\hline
\multicolumn{3}{@{}c}{$\bolds{l=+1}$}& \multicolumn{3}{c}{$\bolds{l=0}$} &\multicolumn{3}{c}{$\bolds{l=-1}$}\\[-6pt]
\multicolumn{3}{@{}c}{\hrulefill}& \multicolumn{3}{c}{\hrulefill} &\multicolumn{3}{c}{\hrulefill}\\
$\bolds{\xi(E_0)}$ &$\bolds{\xi(E_s)}$ &$\bolds{\xi(E_k)}$&$\bolds{\xi(E_0)}$ &$\bolds{\xi(E_s)}$ &$\bolds{\xi(E_k)}$
&$\bolds{\xi(E_0)}$ &$\bolds{\xi(E_s)}$ &$\bolds{\xi(E_k)}$\\
\hline
$\frac{1}{5}\cdot\frac{q+2}{2q+1}$ &$\frac{3}{5}\cdot\frac{3q+1}{2q+1}$ &$0$&$\frac{1}{5}\cdot\frac{q+1}{2q}$ &$\frac{3}{5}\cdot\frac{3q-1}{2q}$ &$\frac{1}{5q}$&$\frac{1}{5}\cdot\frac{q}{2q-1}$ &$\frac{1}{5}\cdot\frac
{(3q-1)(3q-2)}{q(2q-1)}$ &$\frac{2}{5q}$ \\
\hline
\end{tabular*}
\end{table}

%
%
\begin{rem}
Based on our numerical results, we found a remarkable
structure for the $E$-optimal designs with a minimal number of support points
for the second-order response surface model with $k$ predictors,
whenever $k \neq3$. The \mbox{$E$-}optimal design for the second-order
response surface model with a minimal number of support points is
always supported at the sets $E_0$ and $E_k$ and a third set $E_s$. If
$k=3q+l$ where $l=0,\pm1$, then $s=2q+l$. The particular structure is
displayed in Table~\ref{table32}, which also contains the weights
assigned by the $E$-optimal design to these sets.
\end{rem}

%
\begin{exmpl} \label{k=5}
\citet{galil1977b} presented in Table~2 of their paper
$E$-optimal designs (obtained as limits of $\Phi_p$-optimal designs as
$p \to-\infty$).
Note that
not all designs in this class have the minimal number of support points.
For example, if $k=6$ the $E$-optimal design obtained by \citet
{galil1977b} puts masses
$0.040$, $ 0.400$ and $0.560$ at the sets $E_0$, $E_2$ and $E_5$,
respectively, and has
$316$ support points. The $E$-optimal design obtained from Table~\ref
{table32} puts masses
$\xi(E_0)=0.15$, $\xi(E_4)=0.75$,
$\xi(E_6)=0.10 $ and has only $125$ support points.
\end{exmpl}

\section{$E$-optimal designs on the unit ball}\label{sect4}

In this section, we consider the \mbox{$E$-}optimal design problem for the
second-order response surface model
on the $k$-dimensional ball
$ \mathcal{B}_2(1)=\{x\in\mathbb{R}^k\colon\| x \|_2 \leq1\}$.
The general strategy for the solution of the optimal design problem
will be similar as the one given for the cube and we start identifying
a good candidate for the $E$-optimal design.
If the design space is the ball, then
the sets $E_{r_i}$ of barycenters of depth $r_i$
will be replaced by three sets $F_0$, $F_{k-1}$ and $F_k$
as candidate\vspace*{1pt} sets for the support of $E$-optimal designs. Here, $F_0$
consists of the $2^k$ vertices\vspace*{1pt} $x= (\pm
\frac{1}{\sqrt{k}},\ldots, \pm\frac{1}{\sqrt{k}} )^T\in\mathbb
{R}^k$ of the cube $\mathcal{B}_\infty(1/\sqrt{k})$ inscribed in
$k$-dimensional ball $\mathcal{B}_2(1)$,
$F_{k-1}$ consists of the centers $\pm e_i$ of the $(k-1)$-dimensional
faces of $\mathcal{B}_\infty(1)$ [here $e_i=(0,\ldots,0,1,0,\ldots,0)^T$ denotes the $i$th unit vector] and
$F_k$ contains only the center of the ball. Note that the cardinality
of these sets are given by
%
%
\begin{equation}
\label{powersetsofdesign} |F_0|=2^k, \qquad|F_{k-1}|=2k,
\qquad|F_k|=1.
\end{equation}
As a consequence, there is no necessity to search for the minimally
supported designs on the unit ball.

Consider a symmetric design $\xi$ which is supported on the sets
$F_0$, $F_ {k-1} $ and $F_k $ introduced in the previous paragraph. Its
information matrix $M(\xi)$ in the second-order response surface model
(\ref{unconditionedregression}) is of the form (\ref
{informationmatrixforquadraticmodel}) with corresponding eigenvalues
given by (\ref{eigenvaluesab})
where $D= [1-(c-b)-kb]^2+4ka^2>0$. Moreover, from the definition of
$\xi$ we have for the entries defined in the matrix (\ref
{informationmatrixforquadraticmodel})
%
%
\begin{eqnarray}\label{momentsabconball}
a&=&k^{-1}\xi(F_0)+ k^{-1}\xi(F_{k-1}),\nonumber
\\
b&=&k^{-2}\xi(F_0),
\\
c&=&k^{-2}\xi(F_0)+k^{-1} \xi(F_{k-1}),\nonumber
\end{eqnarray}
and it now follows that
%
%
\begin{equation}
\label{identitybetweenEkandaandb} \xi(F_{k-1})= k(a-kb)=k(c-b).
\end{equation}
Substituting this identity into expression (\ref{eigenvaluesab}) for
$\lambda_1$ yields
%
%
\begin{equation}
\label{lambda1asfunctionofa} \lambda_1=\frac{1+a- \sqrt{(1-a)^2+4ka^2}}{2}.
\end{equation}
Therefore, the problem of determining an $E$-optimal (symmetric) design
in the class of measures supported at the sets
$F_0$, $F_{k-1}$ and $F_k$
reduces to the maximization of [note that $a>b$ because otherwise by
(\ref{inequalityabc}) and (\ref{identitybetweenEkandaandb}) we would
obtain $\xi(F_{k-1})=0$, hence $a=b=c$, which is impossible]
%
%
\begin{equation}
\label{phimin} \lambda_{\min}\bigl(M(\xi)\bigr) =\min \biggl\{
\frac{1+a- \sqrt
{(1-a)^2+4ka^2}}{2}, c-b,b \biggr\},
\end{equation}
where $0 \leq a,b,c \leq1$.
In order to construct a good candidate, say $\xi^*$, for the
\mbox{$E$-}optimal information matrix we assume that for the optimal design
all elements in (\ref{phimin}) are identical, which yields by a
straightforward calculation [observing~(\ref{identitybetweenEkandaandb})] for the elements in the matrix (\ref
{informationmatrixforquadraticmodel})
%
%
\begin{equation}
\label{ultimateabandc} a=\frac{k+1}{k^2+2k+2}, \qquad b=\frac{1}{k^2+2k+2}, \qquad c=
\frac
{2}{k^2+2k+2}.
\end{equation}
In this case,
%
%
\begin{equation}
\label{minimaleigenvaluesforball} \lambda_{\min}\bigl(M\bigl(\xi^*\bigr)\bigr)=
\frac{1}{k^2+2k+2}
\end{equation}
is the minimal eigenvalue of the matrix $M(\xi^*)$ with multiplicity
$s=\frac{k(k+1)}{2}$.
Since this solution has been obtained under the constraint that the
designs is supported at the sets
$F_0$, $F_{k-1}$ and $F_k$ and that
all elements in (\ref{phimin}) are identical, it is not clear that the
resulting information matrix is in fact $E$-optimal. In a second
step, we\vadjust{\goodbreak} establish this optimality. In order to explain the general
principle, we begin with an example.

%
\begin{exmpl} \label{exampleextremalpolynomforballandk=2}
Consider the second-order response surface model with $k=2$
predictors. Thus, we have $m=6$ regression functions, the minimum
eigenvalue is given by
$\lambda_{\min}=\frac{1}{10}$, with multiplicity $s=3$. For
a corresponding orthogonal basis in Theorem~\ref{theoremE-optimalityconditionofMelas}, we choose
\begin{eqnarray*}
q_0 &=& (2,-3,-3,0, 0,0)^T,
\\
q_1 &=& (0,-1,1,0,0,0)^T,
\\
q_2 &=& (0,0,0,0,0,1)^T,
\end{eqnarray*}
which yields $ \|q_0\|^2=22$, $\|q_1\|^2=2$, $ \|q_2\|^2=1 $ and
for the extremal polynomial
%
%
\begin{equation}
\label{extremalMelaspolynomforball}
\qquad d(x,\varepsilon)=\frac{w_0}{\|q_0\|^2} \Biggl(2-3\sum
_{j=1}^2x_j^2
\Biggr)^2+\frac{w_1}{\|q_1\|^2}\bigl(x_1^2-x_2^2
\bigr)^2+\frac{w_2}{\|q_2\|
^2}(x_1x_2)^2.
\end{equation}
The vector of weights $w$ is identified by the condition that
there must be equality in~(\ref{A-matrixofMelassequalandmore2}) for
the support\vspace*{1pt} points of the $E$-optimal design and
the condition $w_0 + w_1 + w_2 =1$. Using\vspace*{2pt} the points $x_{(0)}=(0,0)^T
\in F_0$ and $x_{(1)}=(1,0)^T \in F_1$,
we obtain for the vector $w= (\frac{11}{20}, \frac{3}{20},
\frac{6}{20} )$ and
\[
d(x,\varepsilon)= \frac{1}{10} \Biggl(1-3 \Biggl(\sum
_{i=1}^2x_i^2 \Biggr)
\Biggl(1-\sum_{i=1}^2x_i^2
\Biggr) \Biggr)=\frac
{1}{10}\bigl(1-3 \|x\|^2_2
\bigl(1-\|x\|^2_2\bigr)\bigr).
\]
Obviously, we have for all $x$ with ${\|x\|_2\leq1}$
\[
d(x,\varepsilon)\leq\frac{1}{10} = \lambda_{\min}\bigl(M(\xi)
\bigr),
\]
and\vspace*{1pt} by Theorem~\ref{theoremE-optimalityconditionofMelas} any
design with information matrix of the form (\ref
{informationmatrixforquadraticmodel}) with $a=\frac{3}{10}$, $b=\frac
{1}{10}$, $c=\frac{2}{10}$ is $E$-optimal for the second-order
response surface model on the ball.
\end{exmpl}

The following result provides a similar statement in the general case.
Its proof is complicated and therefore deferred to Appendix~\ref{62}.

%
\begin{theor}\label{{theorem_e_optimalityforball_dette_grigoriev}}
Let $\xi^*$ denote a symmetric design on the ball
${\mathcal B}_2(1)$,
which puts masses
%
%
\begin{eqnarray}\label{optimalmeasuresE0Ek1andEkforball}
\xi(F_0)&=&\frac{k^2}{k^2+2k+2}, \nonumber
\\
\xi(F_{k-1})&=&\frac
{k}{k^2+2k+2},
\\
\xi(F_k)&=& \frac{k+2}{k^2+2k+2}\nonumber
\end{eqnarray}
at the sets $F_0$, $ F_{k-1}$ and $F_k$, respectively,
then $\xi^*$ is $E$-optimal for the second-order response surface
model on the
$k$-dimensional unit ball. Moreover,
the minimal eigenvalue of the matrix $M(\xi^*)$ is given by (\ref
{minimaleigenvaluesforball})
with multiplicity $s=\frac{k(k+1)}{2}$ and the
extremal polynomial in Theorem~\ref{theoremE-optimalityconditionofMelas} can be chosen as
%
%
\begin{equation}
\label{normalizedextremalMelaspolynomforball} d(x,\varepsilon)= \frac{1}{k^2+2k+2} \biggl\{1-
\frac{2(k+1)}{k}\|x\|_2^2 \bigl(1- \|x\|_2
^2 \bigr) \biggr\}.
\end{equation}
\end{theor}

We conclude this section with a brief discussion of rotatable designs,
which are defined as designs
for which the dispersion function $U\dvtx  {\mathcal B}_2(1) \to\mathbb
{R};  x \to U(x,\xi) = f^T(x) M^{-1}(\xi) f(x) $ is invariant
with respect to orthogonal transformations, that is,
%
%
\begin{equation}
\label{definitionrotatabelnydesign} U(x, \xi)=U(O x,\xi) \qquad \forall x \in\mathbb{R}^k,
\end{equation}
whenever $O$ is an orthogonal $k\times k$ matrix. Note that this
property is equivalent to the fact that
the function $U(x, \xi)$ depends only of the radius $\| x\|_2$. The
following result characterizes
the rotatability
of a symmetric design with information matrix of the form (\ref
{informationmatrixforquadraticmodel})
and will be used to investigate if $E$-optimal designs in the class of
all rotatable designs are
also $E$-optimal in the class of all symmetric designs.

%
\begin{lemm}\label{theoremrotatablecondition} Let $\xi$ denote a
symmetric design
on the ball ${\mathcal B}_2(r)$ of radius $r >0 $
with information matrix of the form (\ref
{informationmatrixforquadraticmodel}). Then the
design $\xi$ is rotatable for the second-order response surface model,
if and only if the condition
%
%
\begin{equation}
\label{rotatablecondition} c=3b
\end{equation}
is satisfied. Moreover, the uniform distribution on sphere $\partial
{\mathcal B}_2(r)$
denoted by ${{\mathcal U} (\partial{\mathcal B}_2(r) )}$
defines a rotatable design.
\end{lemm}

\begin{pf}
Let $\xi$ denote a design with information matrix (\ref
{informationmatrixforquadraticmodel}). A simple calculation shows that
the inverse
of the $k\times k$ upper block (\ref{matrixA}) of the matrix
$M(\xi)$ is given by
\[
M_{11}^{-1}(\xi) =\pmatrix{ \varkappa& q1_k^T
\cr
q1_k & G},
\]
where $
\varkappa=({c+b(k-1)})/{Q_0}$, $q= {-a}/{Q_0}$,
$Q_0=c-b+(b-a^2)k$, and $G= (d-e) I_k + e 1_k1_k^T$ is a circulant
matrix with diagonal elements $d$ and off-diagonal elements
$e$ defined by
\[
e=\frac{a^2-b}{(c-b)Q_0}, \qquad d=Q_0^{-1}-e(k-1),
\]
respectively.
As a consequence, we obtain for the function $U$ the representation
\begin{eqnarray*}
U(x,\xi) &=& f^T(x)M^{-1}(\xi) f(x)
\\
&=& \varkappa
+\bigl(a^{-1}+2q\bigr)\|x\|_2^2+
\bigl(b^{-1}+2e\bigr)\sum_{i<j}^k(x_ix_j)^2+d
\sum_{i=1}^kx_i^4
\\
&=& \varkappa+\bigl(a^{-1}+2q\bigr) \|x\|_2^2
+ \biggl(\frac{1}{2b}+e \biggr)\|x\|_2^ 4+
\biggl(d-e-\frac{1}{2b} \biggr)\sum_{i=1}^kx_i^4.
\end{eqnarray*}
Now the design is rotatable if and only if the function $U(x,\xi) $
depends only on the radius $\|x\|_2$, that is,
\[
0= d-e-(2b)^{-1}=(3b-c)/2b(c-b), %
\]
which proves the first part of the assertion. The second part follows
by a straightforward calculation of the moments of the uniform
distribution on the
sphere $\partial\mathcal{B}_2(r)$.
\end{pf}

\citet{galil1977a} have determined the $E$-optimal rotatable
designs on
the ball ${\mathcal B}_2(r) $ for the second-order response surface model
(\ref{quadsurf}), which are given by
\[
\xi^* (\alpha)=(1-\alpha)\xi\bigl(\{0\} \bigr)+\alpha{{\mathcal U} \bigl(
\partial{\mathcal B}_2(r) \bigr)},
\]
where the parameter $\alpha$
is defined by
%
%
\begin{equation}
\label{Galil39} \alpha=\cases{ \displaystyle\frac{k(k+1)(k+2)}{(k+1)r^4+k(k+2)^2}, &
\quad$r^2\leq k+2$,
\vspace*{5pt}\cr
\displaystyle\frac{k(r^2-1)}{r^2(r^2+k-1)}, &
\quad$r^2\geq k+2$.}
\end{equation}
If the design space is given by the unit ball ${\mathcal B}_2(1) $ this
design is not $E$-optimal in the class of all designs. In fact, the
symmetric $E$-optimal
design $\xi^*$ determined in Theorem~\ref{{theorem_e_optimalityforball_dette_grigoriev}}
does not satisfy condition (\ref{rotatablecondition}) and is therefore
not rotatable. The minimum eigenvalue of the matrix $M(\xi^*)$ is
given by (\ref{minimaleigenvaluesforball}), while the minimum
eigenvalue of the $E$-optimal design in the class of all rotatable
designs is given by
\[
\lambda_{\min}\bigl( M\bigl(\xi(\alpha)\bigr)\bigr)=\frac{k+1}{k^3+4k^2+5k+1}
< \frac{1}{k^2+2k+2}=\lambda_{\min}\bigl(M\bigl(\xi^*\bigr)\bigr).
\]
We finally note that there exists a difference between the $E$- and
$D$-optimality criterion with respect to the property of rotatability.
In contrast to the $E$-optimal design,
the
$D$-optimal design for the second-order response surface model on the
ball ${\mathcal B}_2(1) $
is also rotatable [see \citet{kiefer1960}].

\begin{appendix}
\section*{Appendix: Proofs of Theorems \texorpdfstring{\lowercase{\protect\ref{theorem_e_optimalityoncubeof_dette_grigoriev}}}{3.1} and \texorpdfstring{\lowercase{\protect\ref{{theorem_e_optimalityforball_dette_grigoriev}}}}{4.1}} \label{sect6}

\subsection{Proof of Theorem \texorpdfstring{\protect\ref{theorem_e_optimalityoncubeof_dette_grigoriev}}{3.1}}\label{61}
\setcounter{equation}{0}

Throughout the proof, we assume
$k\geq2$, the case $k=1$ has been treated in \citet{pukstu1993},
for example.
Recall the definition of the vector of regression functions (\ref
{basicfunctionofquadraticmodel})
in model (\ref{unconditionedregression})
%
and note that for the optimal design $\xi^*$ under consideration we have
$a={2\over5} $ and $b={1\over5}$ in the matrix~(\ref{informationmatrixforquadraticmodel}) with minimum
eigenvalue given by $ \lambda_ {\min} (M(\xi^*)) = \frac{1} {5} $
(see the discussion at the beginning of Section~\ref{sect3}).
Consequently, a possible candidate $q_0, \ldots, q _ {s-1} $ for the
basis of the
eigenspace corresponding to $ \lambda_ {\min} (M(\xi^*)) $ is given by
%
%
\begin{eqnarray}\label{Qmatrix}
Q &=& (q_0, \ldots, q _ {s-1})
\nonumber\\ \\[-16pt]
&=& \pmatrix{ G
_ {k\times(k+1)} &0 _ {k\times k} & 0 _ {k\times ((k (k-1))/2)}
\vspace*{3pt}\cr
0 _ {((k (k-1))/{2}) \times(k+1)} &0
_ {((k (k-1))/ {2}) \times k} & I_ {(k (k-1))/ {2}}}^T,\nonumber
\end{eqnarray}
with an appropriate matrix $ G _ {k\times(k+1)} \in\mathbb{R}^{k
\times k+1} $ (here and throughout this section $0_{r \times s}$
denotes the matrix with all entries given by $0$). This
means that
the unit vectors $e_i=(0,\ldots,0,1,0,\ldots, 0)^T $ are
eigenvectors of the matrix $M(\xi^*) $
for $i=2k+2, \ldots, m={(k+1)(k+2)\over2}$. It turns out that it is
reasonable to use a
vector of weights, which is of the form
%
%
\begin{equation}
\label{optweight} w = (w_0, w_1,\ldots, w
_ {k-1}, 0,\ldots, 0)^T \in\mathbb{R}^s
\end{equation}
in Theorem~\ref{theoremE-optimalityconditionofMelas}. Observing (\ref
{Qmatrix}), it then follows that for vectors of this type only the
$k+1$ functions
\[
\bigl\{ 1, x_1^2, \ldots, x_k^2
\bigr\} %
\]
will appear in the corresponding extremal polynomial.
We now construct the remaining part of the orthogonal basis in (\ref
{Qmatrix}) by choosing the block matrix
%
%
\begin{equation}
\label{ortogonalbasic} G_{k \times(k+1)} =\pmatrix{ k & -2 & -2 1^T_{k-1}
\vspace*{2pt}\cr
0_{k-1} & -1_{k-1} & L}\in\mathbb{R}^{k\times k+1},
\end{equation}
where the matrix $L = (L_{ij})^{k-1}_{i,j=1} \in\mathbb{R} ^ {(k-1)
\times(k-1)} $ is defined by
\[
L _ {ij} = \cases{ -1, &\quad$i+j <k$,
\vspace*{2pt}\cr
k-i, &\quad$i+j=k$,
\vspace*{2pt}\cr
0,
&\quad$i+j> k$.}
\]
This gives for the eigenvectors $q_0,\ldots,q_{k-1}$ [defined by the
first $k$ rows of the matrix $Q$ in (\ref{Qmatrix})]
\[
\|q_0 \| ^ 2=k^2+4k, \qquad
\|q_r \| ^ 2 = (k-r) (k-r+1), \qquad r=1,\ldots, k-1.
\]

With the notation $b_i(x)=(q_i^T f(x))^2$, the extremal polynomial
in Theorem~\ref{theoremE-optimalityconditionofMelas} has the representation
%
%
\begin{equation}
\label{extremalMelaspolynom} d (x, \varepsilon) = \sum_ {i=0}^{k-1}
w_ib_i (x),
\end{equation}
where we have used (\ref{optweight}) and the function $b_0,\ldots,b_{k-1}$ are given by
%
%
\begin{eqnarray}
\qquad b_0 (x) & = & \Biggl(k-2\sum
_ {i=1} ^kx_i^2 \Biggr)
^2\cdot\frac{1}{
\|q_0 \| ^ 2},
\nonumber\\[-8pt]\\[-8pt]
b_r (x) & = & \Biggl(\sum_ {i=1}
^ {k-r} x_i^2 - (k-r) x _ {k-r+1}
^2 \Biggr) ^2\cdot\frac{1} {\|q_r \| ^ 2}, \qquad r=1,\ldots,k-1.\nonumber
\end{eqnarray}

The\vspace*{2pt} coefficients $w_i $ in the
polynomial (\ref{extremalMelaspolynom}) are now
determined by the condition $d(x,\xi)=\lambda_{\min}(M(\xi
))=\frac{1}{5}$ at the points $x^{(r)}=(0,\ldots,0,1,\ldots,1)^T$
with $\| x^{(r)}\|_1=r$ and the fact that $\sum^{k-1}_{i=0}w_i=1$.
This leads to the
matrix equation
%
%
\begin{equation}
\label{systemequation} B(w_0,\ldots,w_{k-1})^T=J_0,
\end{equation}
where
$J_0 = (\frac{1} {5}, \ldots, \frac{1} {5}, 1 )^T \in\mathbb
{R}^{k} $ and
the matrix $B = (B_{ir})_{i, r=0}^{k-1, k-1}$ is a lower triangular
matrix with nonvanishing elements
\[
B_{ir}=\cases{ \displaystyle\frac{k} {k+4}, &\quad$i=0$, $r=0$,
\vspace*{5pt}\cr
\displaystyle\frac{(k-2i) ^2} {k^2+4k}, &\quad$i=1,\ldots, k-2$, $r=0$,
\vspace*{5pt}\cr
\displaystyle\frac{(k-i) ^2} {(k-r) (k-r+1)}, &\quad$i=1,\ldots, k-2$, $r=1,\ldots, i$,
\vspace*{5pt}\cr
1,&\quad$i=k-1$, $r=0,\ldots, k-1$.}
\]
A simple calculation shows
$
(w_0,\ldots,w_{k-1})^T=B^{-1} J_0 =\frac{1} {5k} (k+4, 4,4,\ldots,
4) \in\mathbb{R}^k
$
and $w=(\frac{k+4}{5k}, \frac{4}{5k},\ldots,\frac{4}{5k},0,\ldots,0)^T$ is the vector which will be used for the calculation of a
candidate for the extremal polynomial. For this purpose, we introduce
the notation
%
%
\begin{equation}
\label{valuesofalphar} \alpha_r =\frac{w_r} {\|q_r \| ^ 2} = \cases{
\displaystyle\frac{1} {5k^2}, &\quad$r=0$,
\vspace*{3pt}\cr
\displaystyle
\frac{4} {5k (k-r+1) (k-r)}, &\quad$r=1,\ldots, k-1$,}
\end{equation}
and a tedious but straightforward algebra yields for the polynomial
(\ref{extremalMelaspolynom}) the representation
\begin{eqnarray*}
d (x, \varepsilon) &=& \alpha_0 \Biggl(k-2\sum
_ {i=1} ^kx_i^2
\Biggr)^2 + \sum_ {r=1} ^ {k-1}
\alpha_r \Biggl(\sum_ {i=1}
^ {k-r} x_i^2 - (k-r) x_ {k-r+1}
^2 \Biggr) ^2
\\
&=&\frac{1}{5} \Biggl(1-\frac{4} {k} \sum
_ {i=1} ^kx_i^2
\bigl(1-x_i^2\bigr) \Biggr),
\nonumber
\end{eqnarray*}
which coincides with (\ref{extremalMelaspolynomfactor5}). As a
consequence, we obtain for all $x \in[-1,1]^k$
\[
d\bigl(x,\xi^*\bigr) \leq\lambda_{\min} \bigl(M\bigl(\xi^*\bigr)\bigr) =\tfrac{1}{5}, %
\]
and by Theorem~\ref{theoremE-optimalityconditionofMelas} the matrix
$M(\xi^*)$ is an $E$-optimal information matrix.

\subsection{Proof of Theorem \texorpdfstring{\protect\ref{{theorem_e_optimalityforball_dette_grigoriev}}}{4.1}}\label{62}

The proof proceeds in a similar way as the proof of Theorem~\ref{theorem_e_optimalityoncubeof_dette_grigoriev} but differs in some
essential details from it. To be precise, recall
that for the design $\xi^*$ under consideration the minimal eigenvalue
of its information matrix $M(\xi^*)$ is given by $\lambda_{\min}(
M(\xi^*)) = \frac{1}{k^2+2k+2}$ and has
multiplicity $s = \frac{k (k+1)}{2}$. As in the proof of Theorem~\ref{theorem_e_optimalityoncubeof_dette_grigoriev} we consider the matrix
defined by~(\ref{Qmatrix})
as a candidate for an orthonormal basis of the
corresponding eigenspace. For the matrix $G _ {k\times(k+1)} \in
\mathbb{R}^{k \times k+1}$,
we now use
%
%
\begin{eqnarray}
G _ {k\times(k+1)} & = & \pmatrix{ k & - (k+1) & - (k+1)
1_{k-1}^T
\vspace*{3pt}\cr
0_{k-1} & 1_{k-1} & L},
\end{eqnarray}
{where} $L=(L_{ij})^{k-1}_{i,j=1}\in\mathbb{R} ^ {(k-1) \times(k-1)}
$ is a lower triangular matrix with nonvanishing elements
%
%
\begin{eqnarray}
L _ {ij} &=& \cases{ -i, &\quad$i=j$,
\vspace*{2pt}\cr
1, &\quad$i>
j$.}\label{firstpartortogonalbasicforball}
\end{eqnarray}
Consequently, we have
\begin{eqnarray*}
\|q_0 \|^2 &=& k^2+k (k+1) ^2,
\nonumber
\\
\|q_r \|^2 &=& r (r+1), \qquad r=1,\ldots, k-1,
\\
\|q_r \| ^ 2&=&1, \qquad r=k,\ldots, s-1,\nonumber
\end{eqnarray*}
and with the notation $b_i (x):=( f^T(x) q_i )^2$ the candidate for
the extremal polynomial
in (\ref{A-matrixofMelassequalandmore2}) has the representation
%
%
\begin{equation}
\label{analyticalexpressionofextremalMelaspolynomforball} d \bigl(x, \xi^*\bigr) = \sum_{i=0}^{s-1}
w_ib_i (x),
\end{equation}
where [recall the definition of the vector $f$ in (\ref
{basicfunctionofquadraticmodel})]
\begin{eqnarray}
b_0 (x) & = & \Biggl(k - (k+1) \sum
_ {i=1} ^kx_i^2 \Biggr)
^2\cdot \frac{1} {\|q_0 \| ^ 2},
\nonumber
\\
b_r (x) & = & \Biggl(\sum_ {i=1}
^rx_i^2-rx _ {r+1} ^2
\Biggr) ^2\cdot \frac{1} {\|q_r \| ^ 2}, \qquad r=1,\ldots, k-1,
\nonumber
\\
b_{k-2+i+j} (x) & = & (x_ix_j) ^2,
\qquad i=1,\ldots,k-1;  j=i+1,\ldots,k
\nonumber
\end{eqnarray}
(note that the eigenvectors corresponding to $b_{k-2+i+j}$ satisfy $\|
q_r\|=1$).

For determination of coordinates of the vector $w$, we use again the
fact that there must be equality in condition (\ref
{A-matrixofMelassequalandmore2}) of Theorem~\ref{theoremE-optimalityconditionofMelas} for the support points of an
$E$-optimal design. For the point $x _ {(0)} =0\in\mathbb{R}^k$, the
condition $d (x_{ {(0)}}, \varepsilon) =\lambda_{\min} (M(\xi
^*))$ and
(\ref{analyticalexpressionofextremalMelaspolynomforball})
then yields
%
%
\begin{equation}
\label{co-ordinatep0forball} w_0 =\frac{k^2+3k+1} {k (k^2+2k+2)}.
\end{equation}

We now try to find a candidate for the remaining weights under the
additional assumption that $p_1: = w_1 =\cdots=w_{k-1}$ and $p_2: =
w_k = \cdots
=w _ {s-1} $. Because the sum of all weights is $1$, this gives the equality
%
%
\begin{equation}
\label{identityforco-ordinatesw} w_0 + (k-1) p_1 +\frac{k (k-1)} {2}
p_2=1.
\end{equation}

Finally, we use one more point
$x_{(1)} = (1,0, \ldots, 0)^T \in F_{k-1}$ in the condition
$d (x _ {(1)}, \xi^*) =\lambda_{\min}(M(\xi^*))$ to obtain the equation
%
%
\begin{equation}
\label{co-ordinatep1forball} w_0+p_1\sum
_ {r=1} ^ {k-1} \|q_r \| ^ {-2} =
\lambda_{\min} \bigl(M\bigl(\xi ^*\bigr)\bigr) = \frac{1}{k^2+2k+2}.
\end{equation}
Since $\sum_ {r=1} ^ {k-1} \|q_r \| ^ {-2} =1-k ^ {-1}$,
we finally obtain from (\ref{co-ordinatep0forball})--(\ref
{co-ordinatep1forball}) for the weights
%
%
\begin{eqnarray}\label{co-ordinatep0p1p2forball}
w_0 &=&\frac{k^2+3k+1} {k (k^2+2k+2)},\nonumber
\\
w_1 &=& \cdots=w_{k-1} =\frac{k+1} {k (k^2+2k+2)},
\\
w_k &=& \cdots= w_{s-1}=\frac{2 (k+1)} {k (k^2+2k+2)}.\nonumber
\end{eqnarray}
Substituting these expressions in
(\ref{analyticalexpressionofextremalMelaspolynomforball})
yields by a straightforward calculation
%
%
\begin{equation}
\label{MelasnecesseryandsufficientconditionofEoptimality} d\bigl(x,\xi^*\bigr)= \frac{1}{k^2+2k+2} \biggl(1 -
\frac{2(k+1)}{k} \| x \| ^2_2 \bigl(1- \| x
\|^2_2\bigr) \biggr)
\end{equation}
as a candidate for the extremal polynomial. Obviously, we have
\[
d \bigl(x,\xi^*\bigr) \leq\frac{1}{k^2+2k+2} = \lambda_{\min} \bigl(M
\bigl(\xi^*\bigr)\bigr) %
\]
for all $x \in\mathcal{B}_2(1)$,
and by Theorem~\ref{theoremE-optimalityconditionofMelas} the
information matrix $M(\xi^*)$ defined in~(\ref{informationmatrixforquadraticmodel}) with moments (\ref
{ultimateabandc}) is $E$-optimal for the second-order response surface
model on the ball.
\end{appendix}

\section*{Acknowledgements}
This work was done during a visit of the
second author (Y.~Grigoriev) at the Department of Mathematics, Ruhr-Universit\"at
Bochum, Germany. The authors would like to thank
M. Stein who typed this manuscript with considerable technical
expertise and F. Pukelsheim for some useful hints regarding related
literature. We are also very grateful to two unknown referees for their
constructive comments on the first version of our paper.
The content is solely the responsibility of the authors
and does not necessarily
represent the official views of the National
Institutes of Health.


%

\printaddresses
\end{document}